\documentstyle[epsfig,longtable]{aipproc}

\def \sun {$_{\scriptscriptstyle \odot}$}
\def \ltaprx {\lower .1ex\hbox{\rlap{\raise .6ex\hbox{\hskip .3ex
        {\ifmmode{\scriptscriptstyle <}\else 
                {$\scriptscriptstyle <$}\fi}}}
        \kern -.4ex{\ifmmode{\scriptscriptstyle \sim}\else 
                {$\scriptscriptstyle\sim$}\fi}}}
\def\gtaprx {\lower .1ex\hbox{\rlap{\raise .6ex\hbox{\hskip .3ex
        {\ifmmode{\scriptscriptstyle >}\else 
                {$\scriptscriptstyle >$}\fi}}}
        \kern -.4ex{\ifmmode{\scriptscriptstyle \sim}\else 
                {$\scriptscriptstyle\sim$}\fi}}}

\def\s{{\rm ~s}}

\def\erg{{\rm ~erg}}

\begin{document}
\title{Collapsars}

\author{Andrew MacFadyen}
\address{Astronomy Department, University of California, Santa Cruz, CA 95064\\}

\maketitle

\begin{abstract}

A variety of stellar explosions powered by black hole accretion are
discussed.  All involve the failure of neutrino energy deposition
to launch a strong supernova explosion.  A key
quantity which determines the type of high energy transient produced
is the ratio of the engine operation time, $\rm t_{engine}$, to the
time for the explosion to break out of the stellar surface, $\rm
t_{bo}$.  Stars with sufficient angular momentum produce collapsars --
black holes accreting rapidly through a disk -- in their centers.
Collapsars can occur in stars with a wide range of radii depending on
the amount of pre-collapse mass loss.  The stellar radius and jet
properties determine the degree of beaming of the explosion.  In some
cases the stellar envelope serves to focus the explosion to narrow
beaming angles.  The baryon loading of various models for classical
GRBs formed in massive stars is examined and the consequences are
explored.  For $\rm t_{engine} > t_{bo}$, highly relativistic outflow
is possible and classical GRBs accompanied by supernovae can be
produced.  In other cases hyper-energetic, asymmetric supernovae are
produced.  Longer GRBs ($t \gtaprx 100 \s$) can be produced by
fallback following a weak neutrino-driven supernova explosion.

\end{abstract}

\section*{Introduction}

Massive stars ($M_{ms} \gtaprx$ 25 M\sun) may not always launch
successful neutrino-driven explosions \cite{Fry99}.  We describe here
the continued evolution of such stars after their cores collapse to
black holes and accrete the surrounding stellar mantle.  For the most
massive stars (M$_{ms}$ $\gtaprx$ 35 M\sun) with sufficient angular
momentum, a collapsar -- a rapidly accreting ($\dot{M} \approx$ 0.1
M\sun \s$^{-1}$) stellar mass black hole -- forms promptly at the center
of a collapsing star \cite{MW99}.  We refer to these as Type-I
collapsars.  Less rapidly accreting black holes can also form over
longer time periods due to the fallback of stellar material which
failed to escape during the initial supernova explosions (Type II
collapsars).  This probably happens for main sequence masses, $M_{ms}
\gtaprx$ 20 M\sun.  Stars with masses below this explode as normal
supernovae and leave behind neutron star remnants.  Collapsars power
jetted explosions by tapping a fraction of the binding energy released
by the accreting star through magnetohydrodynamical processes or
neutrino annihilation, or possibly by extracting some of the black
hole spin energy.  The vast majority of stellar explosions do not make
collapsars, only those which make black holes and have sufficient
angular momentum.  Further, not all collapsars make GRBs.  Only those
which happen in sufficiently small (in radius) stars and manage to
accelerate a fraction of the explosion energy to sufficiently high
Lorentz factor.  Other collapsar explosions may be responsible for
hyper-energetic and asymmetric supernovae like SN1998bw.

Table 1 shows a range of observable phenomena possible from collapsars
in various kinds of stars.  Prompt and delayed black hole formation
can occur in massive stars with a range of radii depending on the
evolutionary state of the star, its metallicity and its membership in
a binary system.  The key difference between the scenarios is the
ratio of the time the engine operates, $t_{\rm engine}$, to the time
the explosion takes to break out of the surface of the star,
$t_{bo}$. The breakout time is $t_{bo} = R_{star}/v_{jet}$ where
$v_{jet}$ is the propagation velocity through the star of the
explosion shock or jet head.  Typical velocities are 50,000 km
s$^{-1}$ \cite{Aloy99}. If the engine operates for a sufficiently long
time to continuously power the jet at its base after the
explosion shock (the jet head) breaks out of the surface of the star,
then highly relativistic outflow can be achieved for a fraction of the
ejecta and a classical GRB can result.  The column of stellar material
pushed ahead of the jet, perhaps a few .01 M\sun, escapes the star and
expands sideways leaving a decreasing amount of material ahead of the
jet.

\begin{table}
\caption{Observable phenomena resulting from collapsars.}
\begin{tabular}{|r||c|c|}\hline
 & prompt BH form. & delayed BH form. \\
 & Type I Collapsar & Type II Collapsar\\ 
\hline\hline
No H env. & $t_{\rm engine} > t_{bo}$ & $t_{\rm engine} > t_{bo}$ \\
Wolf-Rayet * & GRB + TypeIb/c SN & long GRB + TypeIb/c SN \\
\hline

Small H env.&  $t_{\rm engine} < t_{bo}$ & $t_{\rm engine} \gtaprx  t_{bo} $ \\
Blue supergiant &  asymmetric SN hard XRT & long GRB? + Type II SN \\
\hline
Large H env. & $t_{\rm engine} \ll t_{bo}$ & $t_{\rm engine} \ltaprx  t_{bo}$ \\
Red supergiant & Asym. hyper-energ. SNII & Asym. hyper-energ. SNII \\
       & Soft Xray transient &  Soft Xray transient + tail \\
\hline
\end{tabular}
\end{table}


The engine time, $t_{engine}$, is the time the star is able to feed
the black hole at a sufficient rate and depends on the stellar mass
and angular momentum distribution at collapse.  Viscous entropy
generation in the accretion disk and centrifugal bounce can eject
significant fractions of the accreting mass flux in a wind
\cite{MW99}.  This can effect the accretion rate onto the black hole
and can shorten the accretion time by expelling the outer layers of
the star and choking accretion onto the disk.  The wind can also be
important for ejecting radioactive nickel into the explosion.  Until
the star starts exploding, the engine time is given by the collapse
time of the star onto the disk.  It is not given by the initial disk
mass divided by the accretion rate since the disk is simply an
intermediate repository of mass that is coming from the collapsing
stellar envelope.  Helium cores can accrete for tens of seconds to
minutes while the accretion time is longer for any size star if the
star initially explodes then partially reimplodes (Type II collapsar)
\cite{MWH99}.  Not all of this time is available for producing an
explosion since there is an initial collapse phase lasting several
seconds when the disk forms, the deposited energy can be advected into
the hole for several seconds until the density is sufficiently low for
energy input to reverse the infall and drive an explosion, and the
explosion takes several seconds to break out of the stellar surface.
The star can also explode after a sound crossing time due to the
lateral expansion of the jet shock especially for ``hot'' jets or due
to the explosion of the star due to a disk wind.

A key characteristic of the model is the degree of spreading of the
jet as it passes through the stellar envelope.  GRBs may have an
approximately common total energy, $E_{\odot} \approx 10^{52} \erg$,
yet produce a variety of stellar explosions depending on the degree to
which the explosion is focussed into a jet.  Two characteristics of
the explosion can determine the beaming of the jet: their entropy and
their duration.  In particular, explosions like SN1998bw can be
explained by the jet being ``hot'' or ``brief''.

\section*{``Hot'' jets}
``Hot'' jets have large internal pressure compared to their ram
pressure and the ambient stellar pressure.  They expand laterally as
they push through the star and share more of their energy with the
stellar envelope.  While it may be possible for a hot jet to escape
the star and make a GRB, more of the star will participate in the
explosion and the jet will take longer to penetrate the star.
``Cold'' jets on the other hand are capable of penetrating the star
with relatively little sharing of energy with the stellar envelope.
In some cases the star actually compresses the jet and helps to focus
it \cite{MWH99,Aloy99}.  In this case the supernova explosion is
relatively weak and a large fraction of the energy goes into the
narrow jet beam.  These types of explosions can have long accretion
episodes and leave large black hole remnants ($M_{\rm bh} >
5M_{\odot}$) since the lateral expansion of the jet is inefficient at
exploding the star and choking the accretion feeding the hole.

\section*{``Brief'' jets}
If the engine is only on for a short time ($t_{engine} < t_{bo}$), the
power is lost before the jet head reaches the surface of the star even
for a small star like the He and C/O cores thought to be responsible
for SN1998bw \cite{Iwa98,woo99}.(Fig 1, right panel).  In this case
the jet expands quasi-isotropically into the star after its ceases to
be energized at it's base (Fig 1, right panel).  The resulting
explosion is asymmetric since the explosion is initiated in the polar
region.  It could be distinguished from a conventional Type Ib/c
supernova by high expansion velocities and large energy.  SN1998bw may
be an example of this with the weak GRB coming from a small amount of
moderately relativistic material \cite{Iwa98,woo99} interacting with
the CSM.

\begin{figure}[t!]
\begin{center}
\begin{tabular}{cc}
\epsfxsize = 8.5 cm
\epsfbox{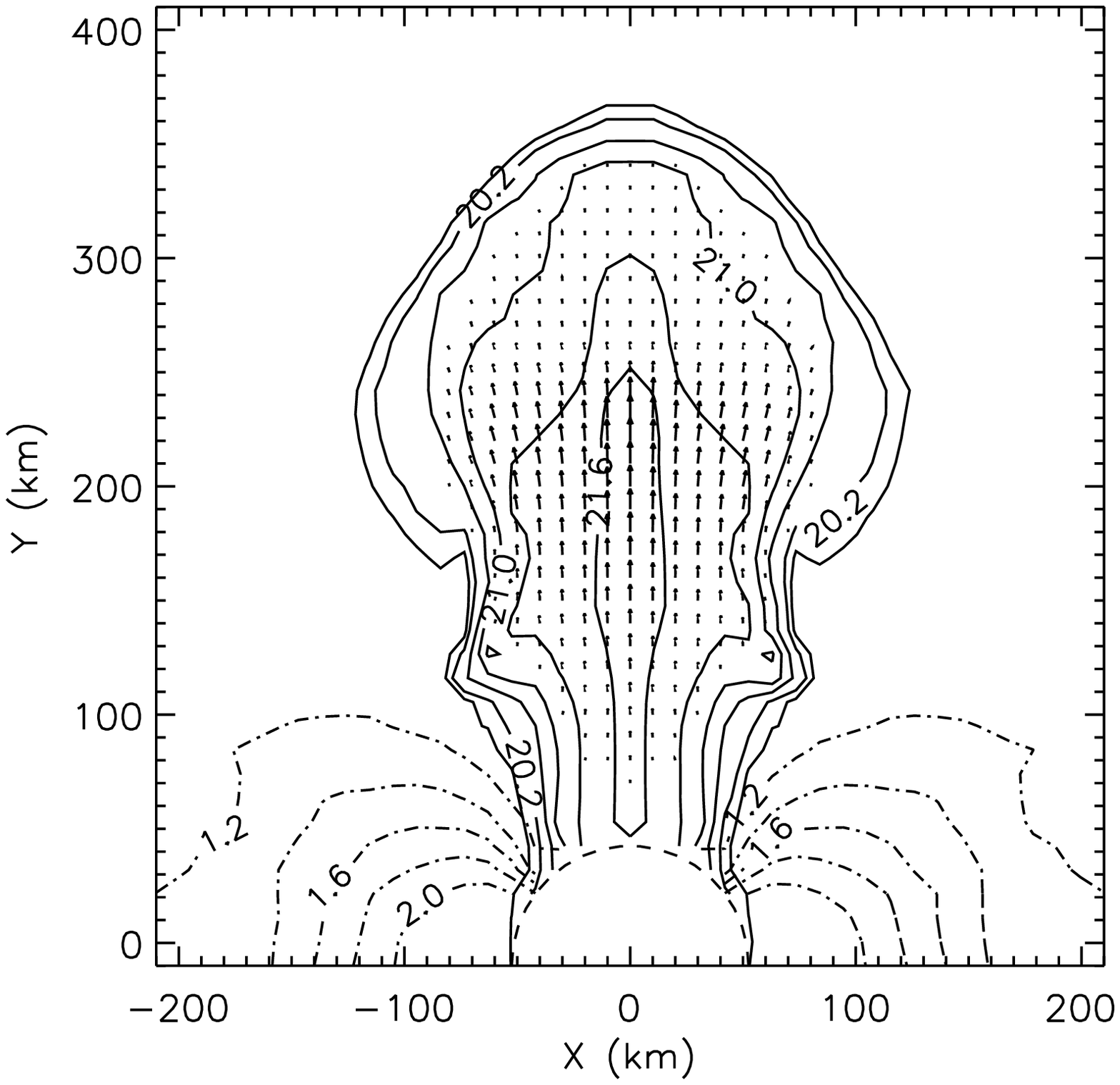}
&
\epsfxsize = 6. cm
\epsfbox{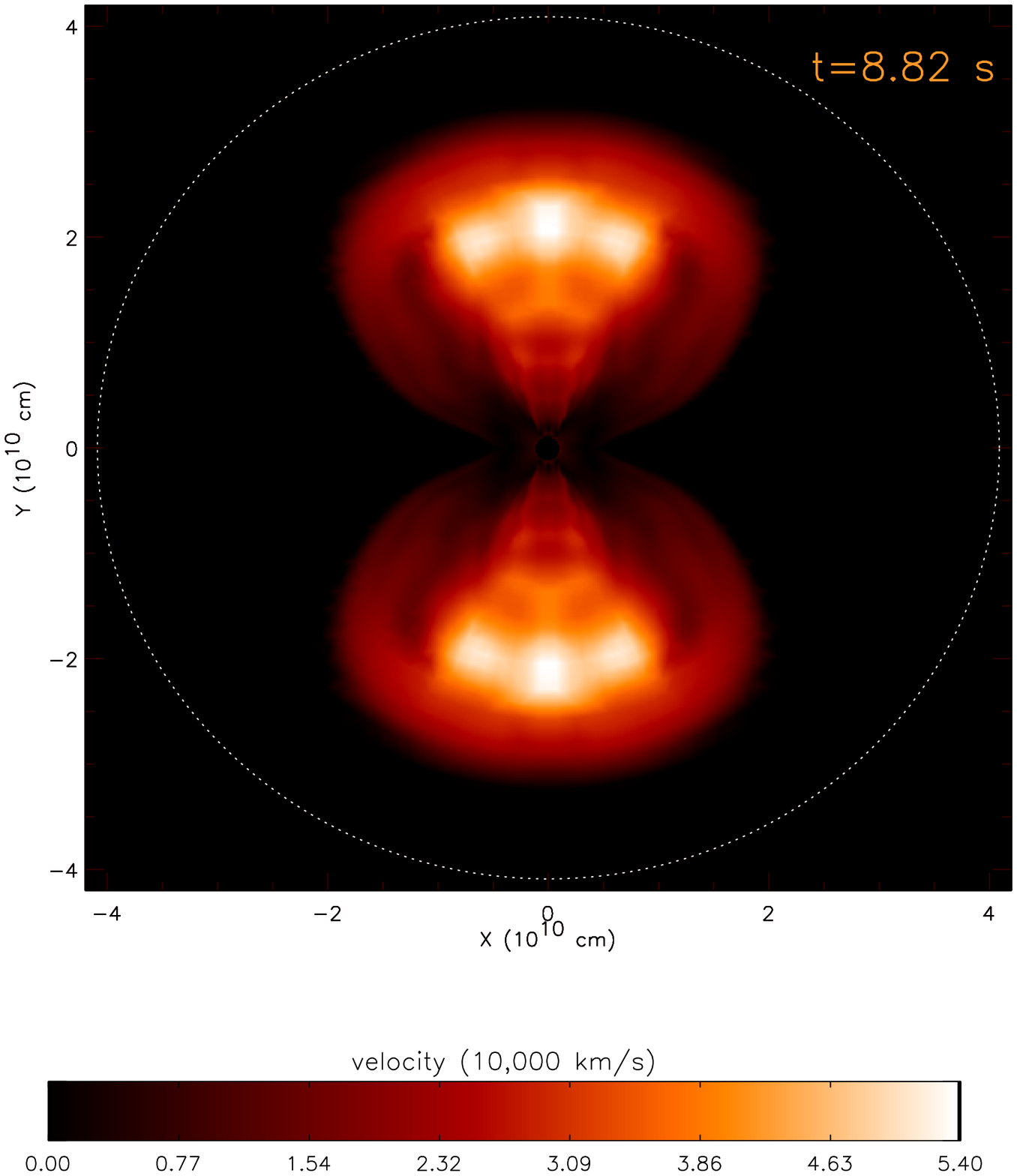}
\end{tabular}
\caption[h]{Left: The solid contour lines show the logarithm of total
(kinetic plus internal) energy density (erg g$^{-1}$) 5.7 milliseconds
after initiating energy input at $5 \times 10^{50}$ erg s$^{-1}$.
After 5.7 ms, 2.8 $\times$ 10$^{48}$ have been deposited and a jet is
starting to form, rising upward along the pole (steepest density
gradient) as shown by the arrows representing the velocity field.  The
magnitude of the largest velocity arrows is $\sim c$ showing that
relativistic expansion has begun.  The dash-dot contours indicate the
temperature in the accretion torus in MeV.  Higher temperature exist
interior to the inner boundary shown here as the dashed semi-circle
with radius 50 km. Right: Asymmetric explosion for $t_{engine} = 5 \s
< t_{bo}$ in a Helium star. $10^{51}$ erg s$^{-1}$ was deposited for
$5 \s$ as thermal energy near the inner boundary in the polar region.
The explosion is shown 3.82 s later..}

\label{fig1}
\end{center}
\end{figure}

\section*{Relativistic Outflow from Massive Stars}

Relativistic outflow can be achieved from an accreting black hole
surrounded by a collapsing stellar mantle.  The key is the prolonged
deposition of energy into regions of the star which can expand due to
their overpressure.  A single impulsive release of energy $E_{dep}(t)=
E_{\circ} \delta(t-t_{\circ})$ will explode a star if the
energy input exceeds the binding energy.  But the explosion will be
``baryon loaded'', a ``dirty fireball'' i.e. a supernova.  The maximum
Lorentz factor will be $\Gamma \ltaprx \frac{E_{dep}}{m_b c^2}$, where
$m_b$ is the mass of baryons in the region where the energy is
deposited.  The Lorentz factor will be less than the asymptotic limit
because the expanding fireball will do work in accelerating the
surrounding material.  {\it For an impulsive energy deposition,} a
clean environment with $m_b < \frac{E_{dep}}{\Gamma c^2}$ and  $\Gamma
\gtaprx 100$ is necessary to make a classical GRB or else the
deposited energy will be shared with too many baryons (the ``baryon
pollution'' problem).

The situation is different if the energy is deposited over a 
period long compared to the expansion time of the energy
loading region.  In this case, energy is injected into a region already
expanding due to the previously deposited energy and the
corresponding overpressure.  Initially, $m_b \gg \frac{E}{\Gamma
c^2}$ and the expansion is sub-relativistic but as the gas expands the
baryon density in the deposition region decreases and the energy per
baryon increases (assuming constant energy deposition rate per unit
volume).  The expanding gas must do work in accelerating its
surroundings so the deposited energy is shared with many baryons and
extremely relativistic motion is initially impossible.  Baryons can be
mixed into the deposition region but centrifugal force and pressure
gradients directed away from the pole can inhibit this (Fig. 1, left
panel).  The amount of baryons which mix into the deposition region is
important for determining the ultimate Lorentz factor that can be
achieved.  Current two-dimensional relativistic calculations for one
particular model achieve $\Gamma \approx 40$ near the deposition
region \cite{Aloy99} and it is expected that higher $\Gamma$ will
result when the calculation is run longer or with greater and/or
variable energy deposition.  Detailed calculations are possible and
will be performed soon.  At late times after the baryons initially
present in the deposition region have expanded away, the Lorentz
factor depends on the energy flux and mass flux into the deposition
region as $\Gamma \approx \frac{\dot{E}}{\dot{m}c^2}$.  The mass flux
depends on the rate that hydrodynamical instabilities mix baryons into
the deposition region and the rate at which the engine injects
baryons.

In the collapsar model for GRBs, or any other similar model involving
a massive star, the key to obtaining relativistic motion is the escape
of an energy loaded bubble from its surroundings (the stellar mantle).
In the case of the toroidal density distribution as in the collapsar
\cite{MW99}, a low density channel is left behind by regions of the
star along the rotational axis which lacked centrifugal support and
fell into the black hole.  Recent simulations \cite{MW99,Aloy99,Mul00}
have shown that energy deposition leads to expansion of gas along the
pole, a jet.  The key to achieving high $\Gamma$ is for energy
deposition to continue at the base of the jet even after the jet head
has broken out of the surface of the star and begun free expansion
into the low density circumstellar environment.  Subsequent energy
deposition at the base of the jet continues to load energy into an
increasingly baryon-free region with the expanding gas continuously
channelled along the rotation axis of the star.

\end{document}